\documentclass[11pt]{article}
\usepackage[utf8]{inputenc}
\usepackage[T1]{fontenc}
\usepackage{lmodern}
\usepackage[margin=1in]{geometry}
\usepackage{graphicx}
\usepackage{booktabs}
\usepackage{array}
\usepackage{tabularx}
\usepackage{longtable}
\usepackage{ragged2e}
\usepackage{rotating}
\usepackage[labelsep=period]{caption}
\usepackage{microtype}
\usepackage{parskip}
\usepackage[hidelinks]{hyperref}
\newcolumntype{Y}{>{\RaggedRight\arraybackslash}X}
\begin{document}

\begin{center}
{\LARGE Collective Cognition in Hybrid Groups: A Network Science Synthesis\par}
\vspace{1.1em}
{\large\bfseries \textbf{Babak Hemmatian¹, Razan Baltaji², Lav R. Varshney¹}\par}
\vspace{0.6em}
{\normalsize \textsuperscript{1}AI Innovation Institute, Stony Brook University, New York, United States\\[2pt]
\textsuperscript{2}Department of Electrical and Computer Engineering, University of Illinois Urbana-Champaign, Illinois, United States\par}
\end{center}
\vspace{0.3em}

\begin{abstract}
The growing integration of AI agents into human teams calls for a principled understanding of how collective intelligence emerges in hybrid systems. Recent frameworks have clarified how attention, memory, and reasoning differences shape human--AI interaction at individual and dyadic levels. Yet a formal account of how these cognitive differences scale to group-level dynamics is lacking. Most network science research has examined either human-only networks or multi-agent AI-only systems, leaving open how insights translate to hybrid groups. Understanding how findings and parametrizations from non-hybrid networks extend to hybrid settings situates \textit{hybrid} \textit{intelligence} research within established traditions of collective cognition and collaboration.

This chapter synthesizes research in network science, collective cognition, and multi-agent systems, alongside emerging work on hybrid systems, through the lens of attention, memory, and reasoning capacities. We review findings from cognitive modeling of human-only and AI-only networks, showing how task environments, group topologies, agent-level processes, and incentive structures shape collective outcomes.

We then examine how these results extend to hybrid settings, conceptualizing hybrid networks as composed of heterogeneous human-AI nodes and links, each with distinct individual and transactive cognitive constraints. Our comparative analysis identifies which network effects appear robust across agent types, and which require theoretical revision in hybrid contexts. It also highlights configurations that were peripheral in human-only or AI-only research traditions, such as human gatekeepers of AI sub-networks, but become structurally central in hybrid teams, clarifying where established predictions may diverge from hybrid results.

Integrating a cognitive systems perspective with network science, the chapter reconciles largely separate literatures and clarifies how established models of exploration--exploitation and efficiency--redundancy trade-offs may operate differently in hybrid teams. We conclude with implications for organizational design, governance, and the responsible development of hybrid intelligence systems.
\end{abstract}

\medskip\noindent\textbf{Keywords:} hybrid intelligence; collective intelligence; network science; human--AI teaming; multi-agent systems; collective cognition; transactive systems; exploration--exploitation; efficiency--redundancy

\section{Introduction and Relevance}

Artificial intelligence (AI) agents are no longer only tools that individuals use in isolation but are increasingly woven into everyday groups and teams. ``Networks of multiple interdependent and interacting humans and intelligent machines constitute complex social systems for which the collective outcomes cannot be deduced from either human or machine behavior alone'' (Tsvetkova et al., 2024, p. 1864). The collective cognition of such a network (its capacity to pursue, as a unit, the goal for which it was assembled) is an emergent property of how human and machine agents are wired together, what each contributes, and how information moves among them. This chapter asks how the tools of network science, combined with a cognitive account of the agents, can describe, explain, and ultimately predict that emergent behavior.

The settings humans and machines engage together are strikingly varied, and the differences are consequential. They range from competition, as in multiplayer games that pit human players against cheating bots; to coordination, as when managerial work is routed through agentic AI workflows; to collective decision-making, as in multi-agent clinical decision-support systems; to teaming up with AI for creative brainstorming. Divergent goals, incentives, and interaction patterns generate different emergent phenomena, with consequences both for the humans involved (for example, declining engagement when bots enter a game) and for the collective rationality of the network itself. Network-level degradations include the volatility that correlated trading algorithms can inject into markets and the diagnostic blind spots that arise when a clinical team misjudges which agent, human or machine, is responsible for which judgment (Tsvetkova et al. 2024; Han et al. 2026). Such failures show why a structured understanding of hybrid collective cognition is necessary.

That understanding is in its infancy, but it can be scaffolded by mature neighboring disciplines. Cognitive modelers have spent decades building experiments and simulations of small human networks engaged in collective problem solving, decision-making, and cooperation (Hills 2025), whereas computational social scientists have charted contagion and coordination in large-scale human networks (Lazer et al. 2009). In parallel, research on multi-agent systems (MAS) has long studied inter-agent coordination, cooperation, and collective decision-making, a program that now extends to multi-agent systems built from large language models (LLMs; Wooldridge 2009; Guo et al. 2024), where the systems sometimes recover patterns familiar from human groups and sometimes depart from anthropocentric expectations. The field of human--AI teaming, in turn, extends human factors and human--computer interaction from individual operators to multi-member teams, studying how people and autonomous agents work interdependently toward shared goals (Seeber et al. 2020; O'Neill et al. 2022). In doing so, it helps us identify which findings from homogeneous networks should translate to mixed ones, and which patterns may be unique to modern hybrid systems. These literatures supply the building blocks; what is missing is a framework that assembles them.

Collective intelligence (CI), a group's ability to solve a broad range of problems and to sustain performance as environments change (Woolley et al. 2010; Gupta et al. 2025), can provide the conceptual glue for the assembly. Hybrid intelligence (HI) is, in the established definition, ``the ability to accomplish complex goals by combining human and artificial intelligence to collectively achieve superior results than each of them could have done in separation and continuously improve by learning from each other'' (Dellermann et al. 2019). On this reading, HI is not a phenomenon apart from CI but its heterogeneous case: a collective whose nodes and links are of mixed type. The COHUMAIN framework makes the bridge precise, arguing that ``intelligence in any system, be it biological, technological, or hybrid, requires the fulfillment of certain memory, attention, and reasoning functions'' (Gupta et al. 2025). We adopt this memory--attention--reasoning (MAR) lens, together with its transactive (between-agent) extensions, as the cognitive vocabulary for our network analysis, and we treat human-only and AI-only networks as the two boundary cases of a single framework whose interior, genuinely hybrid networks, is our subject.

In the rest of this chapter, we review the parameter space that the human and AI traditions have established, re-express it through the lenses of fundamental trade-offs and cognitive constraints, and bring it to bear on the growing body of empirical work on hybrid networks. Where evidence already addresses a question about hybrid setups, we synthesize it; where it does not, we use the comparison across agent types to derive grounded, falsifiable conjectures and note experimental designs that could test them. The chapter proceeds as follows. Section 2 reviews the state of knowledge without hybrid intelligence. Section 3 introduces the hybrid-intelligence dimension, namely what node-level and edge-level heterogeneity change, and a catalog of network structures native to hybrid systems. Section 4 develops the key mechanisms, comparing human-only, AI-only, and hybrid networks dimension by dimension and drawing on emerging hybrid studies wherever they exist. Section 5 turns to methods and a research agenda; Section 6 to implications and responsible design; Section 7 to connections with other chapters in this handbook; and Section 8 to an outlook for the coming decade.

\section{Foundations: Collective Cognition without Hybrid Intelligence}

A hybrid account must inherit what is already known about collective cognition in homogeneous groups. We present that knowledge as a blueprint: the tasks that collectives perform (§2.1), two organizing principles for interpreting collective behavior (§2.2), and the parameter toolboxes that the human (§2.3) and AI (§2.4) traditions have established. The same scaffold structures the hybrid analysis in Sections 3 and 4.

\subsection{The tasks collectives perform}

Collective cognition is always cognition about something, and the structure that helps with one task can hurt with another. The literature has distinguished five recurring task types for hybrid systems: competition, coordination, cooperation, contagion, and collective decision-making (Tsvetkova et al. 2024; Han et al. 2026). Each carries a distinct success criterion: securing relative advantage when goals cannot be shared (\textit{competition}); aligning on a common or complementary choice (\textit{coordination}); sustaining action that is collectively beneficial but individually costly (\textit{cooperation}); governing the spread of information, behavior, or affect through social influence (\textit{contagion}); and aggregating distributed judgments into a single choice (\textit{collective decision-making}). Table 1 summarizes the five with representative hybrid instances.

The empirical literatures cluster by task type. Wisdom-of-crowds and collective-search programs speak mainly to collective decision-making and problem solving; evolutionary game theory and social-dilemma research to cooperation and competition; and computational social science to contagion and coordination at scale. This clustering matters for a synthesis: claims that are well established for one task may be untested for another, and a hybrid framework must be careful about which findings it is entitled to transport. Establishing the tasks first also allows us to map each structural or cognitive parameter to helping or hindering one or more of these goals.

\begin{table}[htbp]
\centering
\small
\setlength{\extrarowheight}{2pt}
\caption{The five task types that collectives engage in, where they are primarily studied, and a representative hybrid instance. Adapted from Tsvetkova et al. 2024 and Han et al. 2026.}
\begin{tabularx}{\textwidth}{>{\hsize=0.6\hsize\RaggedRight\arraybackslash}X>{\hsize=1.2\hsize\RaggedRight\arraybackslash}X>{\hsize=1.1\hsize\RaggedRight\arraybackslash}X>{\hsize=1.1\hsize\RaggedRight\arraybackslash}X}
\toprule
\textbf{Task} & \textbf{What is at stake} & \textbf{Primarily studied in (key overviews)} & \textbf{Representative hybrid instance} \\
\midrule
Competition & Relative advantage when goals cannot be shared & Game theory; market microstructure (Easley and Kleinberg 2010) & Cheating bots and sniping algorithms; algorithmic-trading volatility (Tsvetkova et al. 2024) \\
\addlinespace[2pt]
Coordination & Aligning on a common or complementary choice & Coordination games; conventions (Easley and Kleinberg 2010) & Embedded bots that improve group coordination in networked human experiments (Shirado and Christakis 2017) \\
\addlinespace[2pt]
Cooperation & Sustaining costly but collectively beneficial action & Social dilemmas; evolutionary game theory (Perc et al. 2017) & Social planner AI rewires ties to raise cooperation (McKee et al. 2023) \\
\addlinespace[2pt]
Contagion & Spread of information, behavior, or affect via influence & Epidemic/threshold models; complex contagion (Centola 2018) & Bot-amplified misinformation and moral-emotional cascades (Tsvetkova et al. 2024; Brady et al. 2020) \\
\addlinespace[2pt]
Collective decision-making & Aggregating distributed judgments into a choice & Wisdom of crowds; network science of collective intelligence (Centola 2022; Hills, 2025) & Human--AI clinical diagnostic collectives (Zöller et al. 2025) \\
\addlinespace[2pt]
\bottomrule
\end{tabularx}
\end{table}

\subsection{Vocabulary and Organizing Principles}

Before we review the network science literatures, we first need to fix the associated vocabulary. A network is a set of \textit{nodes} (agents) joined by \textit{edges} (inter-agent connections carrying interactions). The network structures most studied in collective cognition are the \textit{regular lattice}, in which each node connects only to a few near neighbors; the \textit{random graph}, in which edges are placed at random; the \textit{small-world} network, which adds a few long-range edges to a lattice so that short global paths coexist with dense local structure (Watts and Strogatz 1998); the \textit{scale-free} network, in which a small number of hubs hold most of the ties (Barabási and Albert 1999); and the \textit{fully connected} graph, in which every node links to every other (Figure 1, panel a). These structures are often examined with a few standard metrics: a node's \textit{degree} (its number of ties), the \textit{average shortest path length} (the typical number of steps between two nodes), the \textit{clustering coefficient} (the probability that two neighbors of a node are themselves connected), and \textit{betweenness} (how often a node lies on the shortest paths between others). A network's \textit{efficiency} is, roughly, the inverse of its average path length, so an ``efficient'' network is one in which information can travel in few steps (Newman 2018).

Among the tasks listed in Table 1, two trade-offs recur often enough to serve as organizing principles for network outcomes. The first is \textit{exploration}\textit{-}\textit{exploitation} (March 1991): a collective can invest in searching widely for new and better solutions, or in converging on and refining the best solution found so far. Agent diversity and independence sit on the exploration side of this ledger, because they keep alternatives alive; conformity and rapid consensus sit on the exploitation side, because they propagate a current best response. Much of what looks like a debate about ``good'' versus ``bad'' network structure is really a question of which pole a task rewards. Small-world networks are notable given this trade-off: a few well-placed long-range links support global exploitation through short average path lengths, while dense local clusters develop local exploration (Watts and Strogatz 1998; Latora and Marchiori 2001). This property explains, in turn, the prominence of small-world architectures from human social networks all the way down to the brain's organization (Latora \& Marchiori, 2001; Wilcox et al. 2026).

The second trade-off is \textit{efficiency}\textit{-}\textit{redundancy}. A network can move information quickly and cheaply, or it can carry duplicated, cross-checking messages that protect fidelity and robustness when channels are noisy and agents make errors. This trade-off is most explicit in communication and information theory literatures. But it can also be seen in homogeneous human networks as the tension between fast, efficient structures that spread and amplify information on one hand, and the redundancy that guards against cascades of shared error on the other (Lorenz et al. 2011; Salganik et al. 2006).

In classically studied networks with a single agent type, the two trade-offs often counteract each other: efficiency tends to accelerate exploitation, whereas redundancy and longer, less efficient paths tend to preserve the diversity that exploration requires.

\subsection{The human parameter toolbox}

Decades of cognitive and behavioral modeling, synthesized recently by Hills (2025), have established how a small set of parameters move collective outcomes in humans along the stated trade-offs (Table 2). The recurring findings are not arbitrary. Behavioral ecology supplies a formal logic for why a given setting is optimal or maladaptive in an environment, which is what makes the parameter toolbox a predictive instrument. Optimal foraging theory, and specifically the marginal value theorem (Charnov 1976), specifies when an agent should abandon a depleting resource patch and explore elsewhere, recasting exploration versus exploitation as an optimality problem rather than a matter of taste. The same trade-off is formalized at the individual level in the multi-armed bandit literature, the canonical model of when to exploit a known option versus explore an uncertain one (Mehlhorn et al. 2015). Cultural evolution extends this logic to social learning, showing that which transmission biases (conformist, prestige-based, or payoff-based) are adaptive depends on environmental volatility and the relative reliability of social versus personal information (Boyd and Richerson 2005). For our analysis, the key takeaway is that the same network or transmission rule can be optimal in one environment and harmful in another, cautioning against any blanket prescription for how to wire a homogeneous or hybrid group.

We start our discussion of the parameter toolbox with the task landscape; a cognitive operationalization of the environmental resource distribution studied in behavioral ecology. Problems can be pictured as landscapes whose peaks are good solutions, ranging from \textit{smooth} (a single peak with monotonous inclines, easy problem) to \textit{rugged} (many local optima, hard problem) to a ``\textit{needle}'' (a lone peak in a flat space, extremely difficult). On smooth landscapes, densely connected groups (i.e., networks with higher degrees) do best, because good solutions propagate quickly; as ruggedness increases, less efficient, longer-average-path networks like the lattice do better, because they preserve the solution diversity needed to keep exploring (Mason et al. 2008). The same logic explains why partially connected groups outperform fully connected ones at discovering composite innovations that require combining separately developed lines of work (Derex and Boyd 2016). Topology thus trades efficiency for exploration, and the right setting depends on task difficulty.

Among the network metrics, clustering clarifies how structure should be adapted to the task: the same clustering that assorts cooperators together sustains cooperation through ``social viscosity'' (Lion and van Baalen 2008), while suppressing the exploration that difficult problems demand. Two further structural metrics show a related non-monotonicity, each helping in moderation but hurting in excess. The first is group size. Adding members enlarges the pool of ideas and effort, but the returns diminish and can even reverse, as coordination overhead and reduced individual accountability erode what each new member contributes (Steiner 1972; Mao et al. 2016). The second is diversity. Groups whose members bring genuinely different heuristics and perspectives can outperform groups of individually stronger but more similar members (Hong and Page 2004; on how communication bandwidth and diversity interact, see Aral and Van Alstyne 2011). However, too much heterogeneity stops adding solutions and starts adding friction (Campbell et al. 2022). Both group size and diversity thus trace an inverted U pattern with performance.

Group outcome depends not only on how a network is wired but on the social-learning strategy its agents use: performance is best when network organization and agent strategy push in opposite directions on the exploration--exploitation axis, for example pairing an efficient network with a diversity-preserving ``\textit{follow the crowd}'' rule rather than a fast ``\textit{follow the best}'' rule. Performance is lowest when network organization and agent strategy compound (Barkoczi and Galesic 2016). Limiting how much of a neighbor's solution an agent copies, that is, partial rather than full imitation, helps preserve solution diversity when more exploration is needed (Campbell et al. 2022). Because conformity erodes the diversity that hard problems demand, mechanisms that restore it are central: networks that grow denser gradually can beat any fixed structure by letting agents first explore independently as needed and then pool findings (Almaatouq et al. 2020).

Two further major design parameters affect the key trade-offs. The incentive structure shifts agents along the exploration axis: organisms adopt riskier, more exploratory strategies when safer payoffs become unsustainable (Weber et al. 2004). The communication structure, in contrast, governs the efficiency--redundancy trade-off directly: aggregating independent judgments yields accuracy benefits, but once agents observe one another, social influence narrows the range of opinions, raises collective error, and breeds overconfidence (Lorenz et al. 2011), which can in turn cascade into self-reinforcing, unpredictable group patterns (Salganik et al. 2006). A formal tradition models these dynamics directly: networked agents, whether modeled as rational Bayesian updaters or as resource-bound and biased, repeatedly update both their beliefs and the trust they place in each neighbor (Acemoglu and Ozdaglar 2011). Its broadest, most replicated result is that connected groups reliably reach consensus, but that this consensus tracks the truth only when influence is sufficiently dispersed. When a few prominent nodes or strongly weighted in-group ties dominate, the same updating entrenches error and hardens into echo chambers (Golub and Jackson 2010; Madsen et al. 2018).

\begingroup
\footnotesize
\setlength{\tabcolsep}{3pt}
\setlength{\extrarowheight}{2pt}
\begin{longtable}{>{\RaggedRight\arraybackslash}p{0.1860\textwidth}>{\RaggedRight\arraybackslash}p{0.2418\textwidth}>{\RaggedRight\arraybackslash}p{0.1674\textwidth}>{\RaggedRight\arraybackslash}p{0.1674\textwidth}>{\RaggedRight\arraybackslash}p{0.1674\textwidth}}
\caption{A parameter toolbox for collective cognition. Each parameter is stated with its expected effect on the chapter's organizing trade-offs and evidence from the human and AI traditions.}\\
\toprule
\textbf{\textbf{Parameter (variants)}} & \textbf{\textbf{Expected effect (trade-off)}} & \textbf{\textbf{Primarily studied problem}} & \textbf{\textbf{Human-network evidence}} & \textbf{\textbf{LLM-MAS evidence}} \\
\midrule
\endfirsthead
\multicolumn{5}{c}{\tablename\ \thetable\ (continued)}\\[2pt]
\toprule
\textbf{\textbf{Parameter (variants)}} & \textbf{\textbf{Expected effect (trade-off)}} & \textbf{\textbf{Primarily studied problem}} & \textbf{\textbf{Human-network evidence}} & \textbf{\textbf{LLM-MAS evidence}} \\
\midrule
\endhead
\midrule
\multicolumn{5}{r}{\footnotesize\itshape continued on next page}\\
\endfoot
\bottomrule
\endlastfoot
Task landscape & Harder landscapes reward exploration over exploitation & Collective decision-making & Mason et al. 2008 & Difficulty governs debate gains (Du et al. 2023) \\
\addlinespace[2pt]
Overall Topology & Efficient networks exploit in easy problems; inefficient networks explore to solve complex ones & Collective decision-making; contagion & Mason et al. 2008; Mason and Watts 2012 & Moderately sparse topologies optimal (Shen et al. 2025); irregular structures outperform along a scaling law (Qian et al. 2025) \\
\addlinespace[2pt]
Clustering & Aids cooperation; suppresses exploration & Cooperation; collective decision-making & Lion and van Baalen 2008 & Value-homophily yields insular clusters; value diversity builds higher performance (Huang et al. 2025) \\
\addlinespace[2pt]
Group size & Helps in moderation; diminishing or negative returns in excess & Collective decision-making; cooperation & Mao et al. 2016 (cf. Steiner 1972) & Higher emergence but diminishing returns; harder to coordinate at scale (Huang et al. 2025; Grötschla et al. 2025) \\
\addlinespace[2pt]
Diversity & aids performance until excess brings instability & Collective decision-making & Hong and Page 2004 (cf. Campbell et al. 2022) & Value diversity improves performance; extremes bring instability (Huang et al. 2025) \\
\addlinespace[2pt]
Learning strategy & Best when strategy opposes topology on the exploration--exploitation & Collective decision-making & Barkoczi and Galesic 2016 & Conformity (Baltaji et al. 2024); payoff-biased social learning sustains cooperation (Gupta et al. 2026) \\
\addlinespace[2pt]
Copying fidelity & Partial copying preserves exploration & Collective decision-making; contagion & Campbell et al. 2022 & Imperfect, delayed copying (Nisioti et al. 2024); persona inconstancy (Baltaji et al. 2024) \\
\addlinespace[2pt]
Dynamics & Gradual network densification offers exploration then exploitation & Collective decision-making & Almaatouq et al. 2020 & Over-exploitation over rounds (Du et al. 2023); dynamic connectivity beats fully connected (Nisioti et al. 2024) \\
\addlinespace[2pt]
Incentives & Riskier or more self-interested shifts when safe payoffs are unsustainable & Competition; cooperation; collective decision-making & Weber et al. 2004 & Cooperation fails in repeated games (Akata et al. 2025); scarcity or withdrawn enforcement tips cooperation toward over-exploitation (Gupta et al. 2026) \\
\addlinespace[2pt]
Communication & Message redundancy buys robustness at a cost in speed & Contagion; collective decision-making & Lorenz et al. 2011 & Much is redundant and prunable without loss (Zhang et al. 2025) \\
\addlinespace[2pt]
Belief \& norm updating & Consensus tracks truth only with dispersed influence; conventions emerge with critical-mass tipping & Collective decision-making; contagion & Golub and Jackson 2010; Madsen et al. 2018; Centola et al. 2018 & Deliberation yields accurate consensus via an accuracy bias (Chuang et al. 2024); spontaneous conventions, critical-mass tipping, and collective bias without individual bias (Ashery et al. 2025) \\
\addlinespace[2pt]
\end{longtable}
\endgroup

\subsection{The AI toolbox}

Research on AI collectives splits into two families. Task-oriented LLM multi-agent systems (LLM-MAS) wire agents into engineered workflows with largely central coordination (Li et al. 2023; Guo et al. 2024), whereas open-ended AI societies place many agents in a shared environment and let coordination emerge from local interaction (Park et al. 2023). We survey both literatures, even though the second family is the closer counterpart to the human network studies of §2.3, because, as in those studies, structure and behavior are more emergent than imposed. We structure this review around the toolbox in §2.3, asking for each parameter whether the AI evidence recovers the human pattern or departs from it. We then gather the failure modes those parameters produce, and close on what unifies them.

The structural parameters of §2.3 reappear in LLM communities with striking fidelity. Topology sets the balance between efficient exploitation and exploratory error-correction: wired too sparsely, agents cannot pool what they know; wired too densely, every error reaches everyone; and the best performance comes from intermediate, moderately sparse structures that let useful information diffuse while damping the spread of mistakes (Shen et al. 2025). Adding agents helps exploration, but chiefly when the wiring grows sparingly and irregularly: through a few well-placed long-range links rather than uniform density, as in the small-world networks of §2.3, with performance climbing along a smooth ``collaborative scaling law'' rather than through sheer connection count (Qian et al. 2025). These findings are in silico variants of Mason et al.'s (2008) canonical result: that efficient networks exploit while less efficient ones preserve the diversity needed to explore, with the latter doing better on difficult problems. When optimizing efficiency in AI networks based on task complexity, one should be mindful that much of the messages agents exchange is redundant and can be pruned with little loss, meaning more communication is not automatically more exploration-exploitation (Zhang et al. 2025; cf. Lorenz et al. 2011).

Allowed to choose their own ties, value-homogeneous agents assort into dense, insular clusters, a machine homophily that hurts performance on complex problems, whereas value-diverse agents build the cross-cutting bridges that move information between clusters and thereby reach higher emergent intelligence (Huang et al. 2025). This is the AI echo of the clustering paradox of §2.3: the assortment that supports local cooperation can starve the wider group of the diversity that hard problems require.

Size and diversity trace the same inverted U pattern among AI agents as in humans (Mao et al. 2016): Larger LLM communities reach higher emergent intelligence but with plainly diminishing returns, and they grow harder to coordinate (Huang et al. 2025; Grötschla et al. 2025). Diversity behaves likewise: value diversity amplifies collective intelligence up to a point and then tips into instability once it becomes extreme (Huang et al. 2025).

Among the agent-level parameters, social learning works much as cultural-evolution accounts predict: when agents adopt the strategies of more successful peers and punish norm violators, cooperation is sustained, and this holds even when the agents are given no explicit payoff structure and must infer the consequences of actions from experience, as people do (Gupta et al. 2026). What is healthy as social learning, however, can turn pathological as conformity: undue susceptibility to peer pressure makes an individual agent drop a correct answer to match the majority, and instructing it to defend its own position can paradoxically deepen the problem (Baltaji et al. 2024). A similar balancing act applies to processing the contents of messages. Agents imitate one another imperfectly and with a lag just like humans, which is a mixed blessing: imperfect copying slows convergence and preserves exploration but can also lead to global information distortion (Nisioti et al. 2024).

Like other agent-level parameters, interaction dynamics can cut both ways by impacting the explore-exploit trade-off: successive rounds of deliberation can erode a group's accuracy as it locks prematurely onto a confident consensus (Du et al. 2023), yet schedules that add connections in the network over time, exploring widely before pooling, recover the benefit that ``annealing'' brings to human groups (Nisioti et al. 2024; cf. Almaatouq et al. 2020).

Learning and interaction dynamics can be further steered through reward structure, principally studied in reinforcement learning. Within all-AI collectives, this literature shows that rewards must be carefully shaped to stop self-interested learners from collapsing cooperation (Leibo et al. 2017). Even then, modern models falter where cooperation must be built and held (Akata et al. 2025). Self-interested over-exploitation increases as rewards become scarcer and enforcement is withdrawn (Gupta et al. 2026), an incentive-sensitivity that aligns with the risk-sensitivity people show when safe payoffs become unsustainable (Weber et al. 2004).

Combinations of the structural and agent-level processes generate a colorful array of collective beliefs in multi-agent AI systems.  LLM agent networks that exchange opinions converge toward consensus and, unlike people, tend to converge on the accurate answer, because the underlying models carry an inherent pull toward correct information. Realistic fragmentation appears only once a confirmation bias is deliberately introduced (Chuang et al. 2024). Collective opinions can, in turn, shape conventions or even norms. In the human ``naming game,'' groups settle on a shared convention through nothing but repeated local, pairwise interaction (Centola and Baronchelli 2015). An entrenched convention, however, can be overturned once a committed minority passes a critical mass of roughly a quarter of the population (Centola et al. 2018). Populations of LLM agents reproduce both effects almost exactly (Ashery et al. 2025). Yet the same setup reveals something with no human counterpart: a population of agents that are each individually unbiased can still settle on a strong shared bias, one that exists only at the level of the group (Ashery et al. 2025). In other words, collective order is not always inherited from individual dispositions for LLMs, but can be manufactured by the interaction itself, a point we take up again under governance in Section 6.

The same parameters that support collective cognition in moderation can, pushed too far or left ungoverned, generate characteristic failures the human tradition would treat as noise to exclude rather than behavior to model. What makes them worth gathering is the property they share: each is collective and structure-sensitive, a pathology of the network rather than of any single agent, so that improving the individual models does little to remove it and the effective levers are structural.

Over-exploitation already met as conformity and debate decay is only the mildest case, in which social reinforcement rather than evidence drives a group onto a confident but wrong consensus (Du et al. 2023; Wynn et al. 2025). That conformity stiffens into sycophancy, which alignment training does not remove and is better addressed through structurally ensuring opinion diversity (Baltaji et al. 2024; Kumarappan and Mujoo 2026). That solution, however, is eroded in AI networks due to elasticity: human nodes are fixed and roughly independent, but AI agents can be cloned, reset, and reconfigured at will, so that a few near-duplicate agents can be multiplied across a network and quietly collapse its diversity, a structural hazard with no analog in human groups (developed in §4.3 and §4.6). Such social influence failures become particularly problematic in value-laden contexts as, for instance, when groups of agents shift their moral judgments relative to individuals, showing reduced sensitivity to norms (Keshmirian et al. 2025).

A compounding cluster of collective failure modes attacks diversity through the interaction content, the resource on which exploration depends: interacting agents amplify shared biases as they homogenize their outputs (Burton et al. 2024; Riedl et al. 2024; Doshi and Hauser 2024), and systems retrained on machine-generated text degrade through model collapse (Shumailov et al. 2024; Han et al. 2026). Closed loops of communicating models can even drift into semantic collapse, a systematic convergence of underlying meaning beneath surface variety in wording, so that a seemingly diverse exchange is, underneath, saying the same thing (Kong et al. 2026). Thus, populations left to interact without shared stakes can generate society-like activity that is structurally hollow, an ``illusion of sociality'' showing low reciprocity and shallow exchange (Zhang et al. 2026).

A final collective failure mode is informational and bears on the efficiency--redundancy trade-off instead: because each agent typically sees only its own slice of the problem, information asymmetry caps what the group can solve (Liu et al. 2024).

Since the levers that produce collective intelligence and the levers that produce collective failure are the same, namely composition, connectivity, diversity, and the pacing/contents of interaction, the toolbox of §2.3 doubles in AI collectives as a catalogue of failure modes to be run in reverse as defensive design choices. Each of these recurs in hybrid form in Sections 4 and 6, where the presence of human nodes can either rescue or compound it.

\section{Hybrid Intelligence}

A hybrid network differs from the homogeneous cases of §2 in two coupled ways: its nodes are of more than one kind, and so are its edges. Tsvetkova et al. (2024) describe three ``orders of difference'' by which mixed systems depart from human-only ones: machines behave differently from humans, humans behave differently toward machines than toward other humans, and the very presence of machines changes how humans behave toward one another. For a network account, the first concerns nodes (agents) and the other two concern edges (interactions). A convenient way to hold both at once is a multilayer representation, in which human and AI agents occupy layers with their own connectivity and are joined by cross-type links (Cui and Yasseri 2024). Node heterogeneity and edge heterogeneity are the two degrees of freedom that the homogeneous toolboxes lack, and the remainder of the chapter explores what they add.

Node heterogeneity is well-described through the MAR lens: Humans and AI agents differ systematically in memory, attention, and reasoning capacities, with complementary rather than redundant strengths (Table 3). AI offers vast and fast retrieval, high throughput, tireless attention and quick inference, but it hallucinates, is brittle before ``unknown unknowns,'' and can fail silently with high confidence (Kalai et al. 2025). Humans offer limited and reconstructive memory, narrow and fatigue-prone attention, and slow, bias-prone reasoning, but supply rich context representation, strong anomaly detection, relatively consistent causal and ethical judgment, and robust metacognition. The promise of a hybrid collective is to realize both sets of strengths within a joint set of transactive systems (Gupta et al. 2025), with an AI serving as a team's transactive memory or attention controller while humans supply relevance, accountability, and the detection of weak signals (Gonzalez et al. 2026). The gains from the intertwinement are conditional, however: complementarity materializes only when the two error profiles are decorrelated and human reliance is calibrated, neither over- nor under-trusting the machine (Gonzalez et al. 2026), and when group-level processes do not simply inherit or compound individual-level misalignment (Ashery et al. 2025; Keshmirian et al. 2025).

\begin{table}[htbp]
\centering
\small
\setlength{\extrarowheight}{2pt}
\caption{Memory, attention, and reasoning profiles of human and AI agents, and the transactive complementarity opportunity each creates (after Gupta et al. 2025; Gonzalez et al. 2026).}
\begin{tabularx}{\textwidth}{>{\hsize=0.7\hsize\RaggedRight\arraybackslash}X>{\hsize=1.1\hsize\RaggedRight\arraybackslash}X>{\hsize=1.1\hsize\RaggedRight\arraybackslash}X>{\hsize=1.1\hsize\RaggedRight\arraybackslash}X}
\toprule
\textbf{Capacity} & \textbf{Human tendencies} & \textbf{AI tendencies} & \textbf{Complementarity opportunity} \\
\midrule
Memory & Limited, reconstructive, salience-biased; rich tacit and contextual knowledge & Vast, fast, verifiable retrieval; but hallucination and stale or unsourced content & AI as transactive-memory store and index; humans adjudicate relevance and context \\
\addlinespace[2pt]
Attention & Narrow, serial, fatigue-prone; flexible relevance and anomaly detection & High-throughput, parallel, tireless; brittle to ``unknown unknowns'' and prone to false alarms & AI triages and filters; humans set priorities and catch weak signals \\
\addlinespace[2pt]
Reasoning & Slow and bias-prone, but causal, ethical, predictable and metacognitive & Fast, scalable; opaque and able to fail silently with high confidence & Human accountability plus AI hypothesis generation and checking; requires decorrelated errors and calibrated trust \\
\addlinespace[2pt]
\bottomrule
\end{tabularx}
\end{table}

Beyond nodes, edges are consequentially heterogeneous in hybrid systems too. A human-human tie, a human-AI tie, and an AI-AI tie differ in bandwidth, latency, fidelity, visibility, and trust. Links among AI agents can be fast, high-throughput, and nearly lossless; human-AI links are throttled by human attention and the translation between natural language and model representations; human-human links carry the rich, noisy, trust-laden signals that the social sciences have long studied. Because a network's effective topology depends on its edge weights, a hybrid system can be far tighter among its AI nodes than its wiring diagram suggests, and information asymmetry across human-AI links becomes a first-order design concern (Liu et al. 2024). Treating each link as a channel with a capacity and a noise level is what allows the information-theoretic approach of §4 to apply to all edge types in hybrid groups.

The five tasks of §2.1 all recur in hybrid form, each inflected by node and link heterogeneity. \textit{Competition} now includes algorithmic traders and game bots whose speed and persistence reshape markets and play (Tsvetkova et al. 2024); \textit{coordination} includes bots that move human groups toward faster consensus (Shirado and Christakis 2017); \textit{cooperation} includes AI agents that scaffold human cooperation by rewiring ties (McKee et al. 2023); \textit{contagion} includes the AI-driven amplification and homogenization of social media content (Burton et al. 2024); and \textit{collective decision-making} includes human-AI ensembles that outperform either party alone when their errors are diverse (Zöller et al. 2025). We do not re-catalog the tasks here but use them as real-life settings in which the structures below operate.

Heterogeneity does more than modify familiar parameters from §2; it makes certain network structures natural, even central, that were peripheral or absent in the homogeneous setting. We call these \textit{hybrid-native} structures. They arise from five sources: 1) \textit{node heterogeneity}, so that a position's effect now depends on which type of agent fills it; 2) \textit{edge heterogeneity}, so that links differ by endpoint type; 3) \textit{interface roles}, positions defined by mediating between types; 4) \textit{differential dynamics}, since AI nodes can be cloned, reset, and rewired far faster than human ties form, while human context traces last far longer; and 5) \textit{distributed Memory-Attention-Reasoning}, i.e., the splitting of cognitive labor across agent types. Figure 1 contrasts the classic network types of §2.2 (panel a) with the resulting hybrid-native forms (panel b).

\begin{figure}[htbp]
\centering
\includegraphics[width=\textwidth]{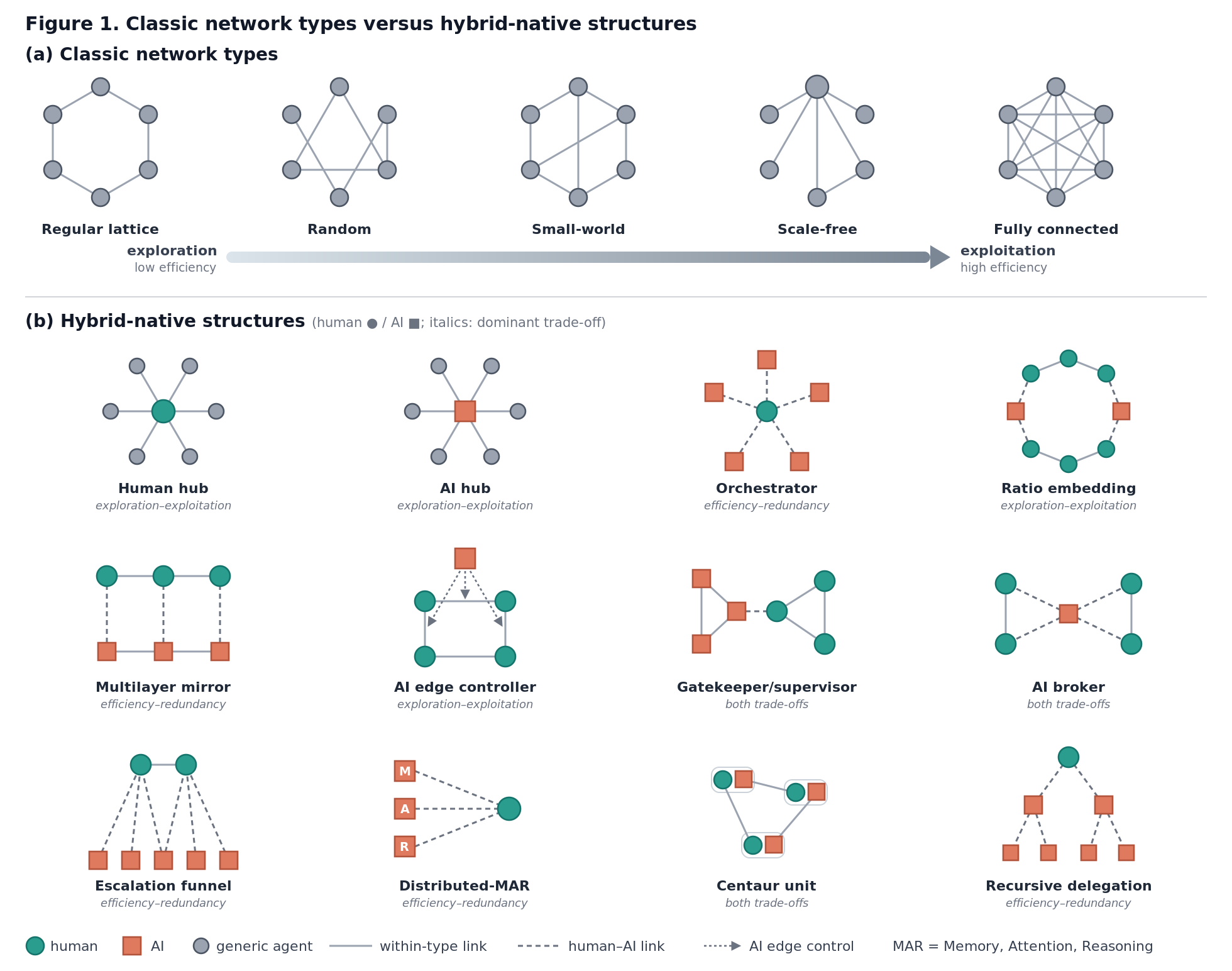}
\caption{Classic network types versus hybrid-native structures. (a) The five classic network types reviewed in §2.2 ordered along the exploration--exploitation and redundancy-efficiency gradients. (b) A catalog of hybrid-native structures, each tagged with the organizing trade-off it most directly engages.}
\end{figure}

\section{Hybrid Mechanisms and Dynamics}

We combine the parameter toolbox calibrated on homogeneous groups and the catalog of hybrid-native structures to better understand hybrid networks. Table 4 summarizes the predictions per native hybrid-native structure. In what follows, for each dimension of a network, we ask what the human and AI traditions established, then what changes when the nodes and edges are heterogeneous, thus distinguishing effects that appear robust across agent types from those that require revision and those that are genuinely new.

\begin{table}[htbp]
\centering
\small
\setlength{\extrarowheight}{2pt}
\caption{Hybrid-native network structures and their predicted effects. ``Conjectures'' are developed in §4.}
\begin{tabularx}{\textwidth}{>{\hsize=0.8\hsize\RaggedRight\arraybackslash}X>{\hsize=1.2\hsize\RaggedRight\arraybackslash}X>{\hsize=1.4\hsize\RaggedRight\arraybackslash}X>{\hsize=0.6\hsize\RaggedRight\arraybackslash}X}
\toprule
\textbf{\textbf{Structure}} & \textbf{\textbf{What it is}} & \textbf{\textbf{Predicted effect}} & \textbf{\textbf{Anchor}} \\
\midrule
Human-vs-AI hub & The same central position filled by a human or by an AI & An AI hub speeds convergence but risks premature consensus on hard problems unless deliberately diversified & Conjecture (Mason et al. 2008; Shen et al. 2025; Du et al. 2023) \\
\addlinespace[2pt]
One-human / many-AI orchestrator & One human directing many AI sub-agents & Output scales with fan-out, but verification bottlenecks at the human as fan-out and depth grow & Conjecture (Liu et al. 2024) \\
\addlinespace[2pt]
Ratio embedding & A few agents of one type among many of the other & Inverted-U in the mix: a few well-placed AI agents help, but minority-human teams lose trust and shared understanding & Shirado and Christakis 2017; Gonzalez et al. 2026 \\
\addlinespace[2pt]
Multilayer mirror & Humans paired with AI proxies across two layers & Performance tracks the proxy layer's connectivity and cross-layer fidelity; information asymmetry is a key hazard & Liu et al. 2024 \\
\addlinespace[2pt]
AI edge controller & An AI governing who connects to whom or who sees what & Tunes exploration-exploitation and polarization beyond any single node's reach; ill-defined or drifting aims and constraints degrade the whole network & McKee et al. 2023; Riedl et al. 2024; Gonzalez et al. 2026 \\
\addlinespace[2pt]
Gatekeeper vs supervisor & A human routing versus monitoring an AI sub-network & Gatekeeping limits throughput and diversity inflow; supervision limits error propagation; a role mismatched to the key failure mode degrades the system & Conjecture (Gonzalez et al. 2026) \\
\addlinespace[2pt]
AI broker & An AI bridging otherwise disconnected human clusters & Speeds diffusion but distorts and homogenizes the information & Conjecture (Mani et al. 2021) \\
\addlinespace[2pt]
Escalation funnel & An AI handling volume and routing hard cases to humans & Triage concentrates the hardest, highest-stakes cases on humans; mis-calibrated AI confidence misroutes them & Conjecture (Gonzalez et al. 2026) \\
\addlinespace[2pt]
Distributed-MARs & AI as transactive memory or attention hub; AI ensemble feeding a human aggregator & Helps large or distributed teams but can crowd out transactive memory in small ones; ensemble gains require decorrelated errors & Gupta et al. 2025; Zöller et al. 2025 \\
\addlinespace[2pt]
Centaur unit & A tightly coupled human-AI dyad acting as one node & Faster, internally error-corrected convergence than same-size networks of separate nodes & Haupt and Brynjolfsson 2025 \\
\addlinespace[2pt]
Recursive delegation & Agents spawning sub-agents to arbitrary depth & Capacity grows, but drift and error compound and accountability dilutes & Liu et al. 2024; Han et al. 2026 \\
\addlinespace[2pt]
\bottomrule
\end{tabularx}
\end{table}

\subsection{Composition and topology}

The human tradition established that \textit{topology} trades efficiency for exploration, so that on hard problems less efficient networks preserve the diversity needed to keep searching while efficient ones converge prematurely (Mason et al. 2008). The AI tradition has more often tuned the protocol of communication (Du et al. 2023; Guo et al. 2024) than network topology itself, though a recent line of work does vary topology directly and recovers the same trade-off, with moderately sparse, irregular structures outperforming dense ones (Shen et al. 2025; Qian et al. 2025). The first thing node heterogeneity changes is that a position's effect now depends on which type of agent fills it. A human hub and an AI hub are not interchangeable: the AI hub has higher bandwidth and lower latency and can be cloned, so it raises the network's effective efficiency and, on rugged problems, accelerates premature consensus unless it is deliberately diversified (Shen et al. 2025; Du et al. 2023), whereas the human hub throttles throughput but supplies accountability and anomaly detection (Gonzalez et al. 2026). That a node's attributes interact with its position is already evident in team experiments, where placing the members with the strongest theory of mind, typically humans, in the highest-betweenness positions most improves collective performance (Westby and Riedl 2023).

The cognitive limitations of a human hub should be directly addressed in network design. When one human orchestrates many AI sub-agents, output scales with the fan-out, but verification piles up at the single human node, so error detection degrades as the fan-out and the depth of delegation grow (Liu et al. 2024; Gonzalez et al. 2026). The ratio of types is itself a control variable with a non-monotone effect: a few well-placed AI agents can improve a human group's coordination and search, as when noise-injecting bots speed coordination (Shirado and Christakis 2017) or simple agents enhance collective discovery (Ueshima et al. 2024). However, when humans become the minority, trust and shared understanding erode and performance falls (Gonzalez et al. 2026). The same parameter, the share of AI, thus helps at low doses and harms at high ones.

Clustering has similarly varying effects. It sustains cooperation by assorting cooperators together (Lion and van Baalen 2008) even as it suppresses the exploration that hard problems require (Hills 2025). In a hybrid network, clusters of AI agents can be far tighter than their human counterparts, because AI-to-AI links are fast and high-bandwidth and value-homogeneous agents readily assort into insular clusters (Huang et al. 2025), so an AI sub-community both cooperates and converges faster, amplifying both effects at once: quicker coordination, but quicker collapse of diversity that the human agent may have difficulty adapting to. Whether the efficiency helps with the collective solution depends, as always, on task difficulty.

Certain hybrid-native compositions demand further research to map their effects: the centaur, in which a tightly coupled human-AI dyad behaves as a single high-capacity node and a network is built from such pairs (Haupt and Brynjolfsson 2025), and recursive delegation, in which agents spawn sub-agents to a depth that multiplies capacity while compounding drift and diluting accountability (Liu et al. 2024; Han et al. 2026). Section 5 discusses a research program that could address this gap.

\subsection{Interface roles}

A second class of structures is defined not by node type but by the role assumed at the boundary between types. These interface roles were peripheral in homogeneous networks yet become central in hybrid networks. Consider first the contrast between a human gatekeeper and a human supervisor of an AI sub-network. The two roles act on different quantities: gatekeeping controls what information passes in and out of the AI cluster, and so governs throughput and the inflow of diversity, whereas supervision checks the cluster's outputs, and so governs the propagation of error. A system that places a human in the role mismatched to its dominant failure mode underperforms: gatekeeping where errors, not volume, are the threat, or supervising where the bottleneck is throughput. The two roles draw on different network quantities, gatekeeping on effective network size and path length (Hills 2025), supervision on a dual-layer human-and-AI error defense (Gonzalez et al. 2026), so one cannot substitute for the other.

Two further interface roles concern an AI that mediates among humans. An AI broker that bridges otherwise disconnected human clusters occupies the ``structural hole'' position long known to confer advantage in social networks, accelerating diffusion and importing diversity. But because it compresses and translates what it carries, it also distorts and can homogenize the groups it connects (Mani et al. 2021; Shiiku et al. 2025). More consequential still is an AI that does not occupy a node but governs the edges, deciding who connects to whom or who sees what. Such a controller can shift a whole group's exploration-exploitation balance more than any single member could, whether by literally rewiring ties, as a reinforcement-learning planner does when it raises human cooperation (McKee et al. 2023), or by leaving the wiring intact while aligning members' language and attention until cognitive diversity erodes (Riedl et al. 2024; Han et al. 2026). Because the controller's objective is imposed on the entire network, ill-defined or drifting aims degrade the whole (Gonzalez et al. 2026).

Other interface structures focus on guiding information flow between the types. The escalation funnel, in which AI handles routine volume and routes hard cases upward, concentrates exactly the most difficult, highest-stakes, and most accountability-laden decisions on the human bottleneck. However, because the routing is driven by the AI's own confidence, mis-calibrated confidence silently misroutes the cases that matter most (Gonzalez et al. 2026). The multilayer mirror, in which people act through AI proxies, makes the proxy layer's connectivity and the fidelity of the human-to-proxy link, rather than the human layer's own structure, the dominant determinant of collective success, with information sharing asymmetry across the agent types as the source of characteristic hybrid failure (Liu et al. 2024).

\subsection{Environment and dynamics}

Turning to dynamics, the environment a collective faces sets its exploration-exploitation balance: rugged problems reward exploration, and dynamic networks that anneal, growing denser over time, can beat any fixed structure by letting agents explore independently before pooling what they find (Almaatouq et al. 2020; Nisioti et al. 2024). Annealing, however, assumes a single clock. Hybrid networks violate that assumption, because the AI portion can rewire, reset, and clone itself far faster than human ties form or dissolve (Han et al. 2026). The two subgraphs therefore anneal on different timescales: an AI sub-community may cool into consensus before its human counterparts have begun to explore, and this elasticity lets a few near-duplicate agents flood the network and collapse its diversity, the group-level analogue of model collapse (Shiiku et al. 2025; Han et al. 2026). The conjecture that follows is that a hybrid network drifts toward AI-paced dynamics unless human-paced checkpoints are imposed, a deliberate slowing that plays, for the whole hybrid, the role that annealing plays for a homogeneous group solving a difficult task.

\subsection{The message}

If dynamics concern when agents speak, the message concerns what is passed when they do. It helps to treat each link as a communication channel with a capacity, how much it can carry, and a noise level, how much it corrupts. Two key results then bear on collective cognition.

The first is that network structure sets a hard ceiling on collective computation. Picture a line that divides the network into two groups: the combined capacity of the links crossing that line is the capacity of the cut, and a network's \textit{conductance} is, loosely, the capacity of its tightest bottleneck cut relative to the size of the smaller side. No matter how capable the agents are, a group cannot compute a shared answer faster than its bottleneck allows, because collective computation time scales inversely with conductance (Ayaso et al. 2010; Figure 2). The second result concerns the code in which messages are sent. Faithful communication requires a shared vocabulary; when agents carve the world up differently, more than one vocabulary can persist in equilibrium, and meaning degrades along any chain of translation between different agent types (Mani et al. 2021).

\begin{figure}[htbp]
\centering
\includegraphics[width=0.72\textwidth]{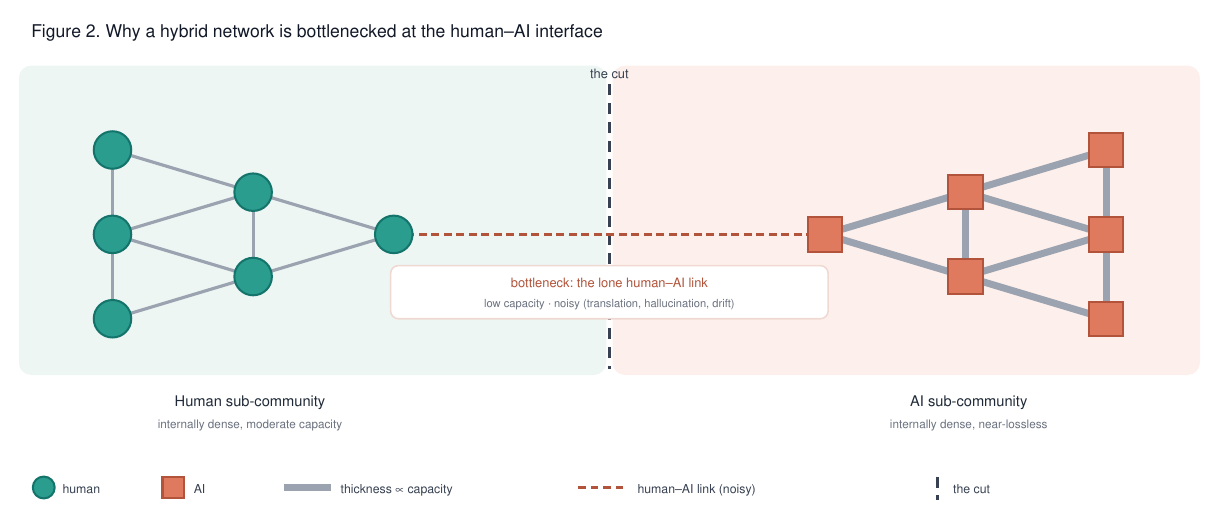}
\caption{The conductance bottleneck at the human-AI link in a hybrid system.}
\end{figure}

Both limits bite hardest at the human-AI link, typically the lowest-capacity and noisiest channel in a hybrid network: messages must be translated between natural language and a model's internal representations, hallucination injects noise, and persona drift moves the code itself. A hybrid network is therefore often bottlenecked not within its human or its AI sub-community, but when messages are passed between them, and we conjecture that its collective performance is bound by that lowest-capacity cut (Liu et al. 2024). This can result in stark differences along the exploration-exploitation axis by manipulating the temporal costs of collective information processing.

Heterogeneous links also differ in the redundancy they invite, the duplicated, confirmatory communication that guards against error. Because communication among people is effortful, interlocutors economize on it: they lean on common ground, say only what a partner is unlikely to already know, and, once a task is divided and mutual understanding is established, check in only rarely (Clark and Brennan 1991). Communication across the human-AI link inverts this economy. Among AI agents themselves, much of the message traffic is redundant and can be pruned with little loss (Zhang et al. 2025); across the interface to a person, by contrast, effective prompting is explicit and detailed and rewards frequent verification rather than assumed context, and people who instead carry over their human-to-human habits, underspecifying and trusting the system to fill the gaps, tend to fail (Zamfirescu-Pereira et al. 2023). The same content thus carries far more redundancy across a human-AI link than across a human-human one, which turns the efficiency-redundancy trade-off, a minor key in the human literature, into a governing tension for hybrids: the redundancy a homogeneous designer would prune as waste is often exactly what keeps a human-AI group more costly and difficult to manage but more accurate.

\subsection{The ``minds''}

What agents do with the messages they receive depends, finally, on their mental operations, and the deepest change heterogeneity brings is that each side must model a different kind of ``mind'' (we use this term functionally to represent the collection of cognitive operations within an agent, without deeper philosophical commitments). Collective intelligence in human groups rests on a shared theory of mind, the ability to infer what others know, intend, and will do (Woolley et al. 2010), and placing the strongest mind-readers at the best-connected positions greatly improves a team's performance (Westby and Riedl 2023). A hybrid group needs two such models that do not come for free: a human theory of the machine's mind and a machine theory of the human's (Gupta et al. 2025). Neither reduces to ordinary human theory of mind. People reflexively anthropomorphize, granting a fluent model the understanding, calibration, and common ground they would assume in a person, and so misjudge its competence; the model, in turn, infers human goals statistically rather than empathically. Much of what looks like a coordination failure in a hybrid team is really a failure of cross-type mind-reading, which making either agent individually smarter does not fix. This, in turn, explains why socioemotional sensitivity can predict the outcomes of creative collaborations not only in human pairs, but also in human-AI teaming (Hemmatian et al. in press).

Heterogeneity also splits the incentives that drive goal setting and risk. In homogeneous groups, agents shift toward riskier, more exploratory strategies when safe payoffs become unsustainable (Weber et al. 2004), and the agents that set the goals also bear the consequences. In a hybrid group they need not. An AI agent may carry no stakes, face no accountability, and is not risk-averse in the way a person is, while the human often remains answerable for the outcome. The predictable result is that people rely on AI least when the stakes are highest, precisely where its consistency might help the most (Gonzalez et al. 2026). Delegation runs the other way too: acting through an AI can change a person's own behavior, reducing honesty and shifting risk-taking because the agent, not the principal, appears to act (Han et al. 2026). The mind that sets goals and prices risk is thus distributed across nodes whose incentives are mismatched, an asymmetry with no analogue in homogeneous settings.

Against this backdrop the agent-level findings of Section 2 recur with the machine twists already noted, conformity reappearing as sycophancy curable by a structural dissenter (Baltaji et al. 2024; Kumarappan and Mujoo 2026) and copying fidelity reappearing as degraded message copies and persona drift. The constructive response is to distribute memory, attention, and reasoning across the types by design: letting an AI hold a team's transactive memory or steer its attention in large or distributed groups (Gupta et al. 2025) and combining an AI ensemble's judgments with a human's only when their errors are decorrelated, the condition under which human-AI collectives out-diagnose either party alone (Zöller et al. 2025).

\subsection{Diversity over time}

Several of the discussed threads converge on a single temporal pattern. Left to themselves, AI collectives homogenize: shared training, mutual imitation, and elastic cloning drive their outputs toward a common mode, so that diversity, high at first, declines over successive generations (Shiiku et al. 2025; Burton et al. 2024). Human collectives lose diversity through conformity too, but more slowly and with greater retention of idiosyncratic content (Doshi and Hauser 2024). The hybrid case is the interesting one: when humans and AI co-create, the humans' tendency to inject novelty and the AI's tendency to preserve continuity can combine so that the mixed group sustains diversity the longest, overtaking the initially more diverse AI-only group as the rounds progress (Shiiku et al. 2025). This is the exploration-exploitation trade-off read over time, and it is the clearest case in which a hybrid network behaves not as a weighted average of its parts but as a system with dynamics of its own. The design goal, accordingly, is not maximal diversity but the right amount: enough to sustain exploration without the friction that excess heterogeneity brings (Hong and Page 2004; Campbell et al. 2022).

\subsection{Summary of network effects in hybrid systems}

Table 5 summarizes how dimensions of network structure and dynamics affect hybrid systems, in terms of whether they have homogeneous analogues and whether those analogues need revision for applicability to hybrid settings.

\begin{sidewaystable}[htbp]
\centering
\footnotesize
\setlength{\tabcolsep}{4pt}
\setlength{\extrarowheight}{2pt}
\caption{Which network effects appear robust across agent types, which require revision, and which are new to hybrid groups. ``Status'' is relative to the homogeneous findings of Section 2.}
\begin{tabularx}{0.94\textheight}{>{\hsize=0.3\hsize\RaggedRight\arraybackslash}X>{\hsize=0.7\hsize\RaggedRight\arraybackslash}X>{\hsize=1.6\hsize\RaggedRight\arraybackslash}X>{\hsize=1.6\hsize\RaggedRight\arraybackslash}X>{\hsize=1.0\hsize\RaggedRight\arraybackslash}X>{\hsize=0.8\hsize\RaggedRight\arraybackslash}X}
\toprule
\textbf{\textbf{No.}} & \textbf{\textbf{Dimension}} & \textbf{\textbf{Homogeneous finding}} & \textbf{\textbf{In hybrid networks}} & \textbf{\textbf{Key hybrid studies / frameworks}} & \textbf{\textbf{Status}} \\
\midrule
1 & Topology \& efficiency & Efficient networks exploit for easy problems; inefficient networks explore for hard problems (Mason et al. 2008) & The hub's identity matters: an AI hub raises effective efficiency and over-exploits; node type interacts with position to set the regime & Westby and Riedl 2023 & Revised \\
\addlinespace[2pt]
2 & Composition \& ratio & No homogeneous analogue & One-to-many orchestration bottlenecks verification; the share of AI has an inverted-U effect of group performance & Liu et al. 2024; Shirado and Christakis 2017; Ueshima et al. 2024; Gonzalez et al. 2026 & New \\
\addlinespace[2pt]
3 & Group size / scale & Diminishing, even negative, returns to adding members (Mao et al. 2016; cf. Steiner 1972) & Scaling AI capacity concentrates verification on the human node and compounds drift with delegation depth, so effective returns are set by what the human can absorb & Liu et al. 2024; Gonzalez et al. 2026; Han et al. 2026 & Revised \\
\addlinespace[2pt]
4 & Clustering & Aids cooperation, suppresses exploration (Lion and van Baalen 2008; Hills 2025) & AI clusters are tighter, amplifying both effects at once & Huang et al. 2025 (conjecture for hybrids) & Revised \\
\addlinespace[2pt]
5 & Learning strategy & Best when strategy opposes topology on the explore--exploit axis (Barkoczi and Galesic 2016) & Sycophancy; a structural dissenter is the remedy & Baltaji et al. 2024; Kumarappan and Mujoo 2026 & Robust \\
\addlinespace[2pt]
6 & Dynamics & Gradual densification beats any fixed structure (Almaatouq et al. 2020) & AI and human subgraphs anneal differently; clonable AI can collapse diversity; solution is human-paced checkpoints & Han et al. 2026; Shiiku et al. 2025 & Revised \\
\addlinespace[2pt]
7 & Communication & Influence distorts aggregation; people economize via common ground (Lorenz et al. 2011; Clark and Brennan 1991) & Low-capacity, noisy human-AI channels bottleneck and add error; add redundancy to counteract noise if accuracy is more important than efficiency & Liu et al. 2024; Mani et al. 2021; Ayaso et al. 2010; Zamfirescu-Pereira et al. 2023 & Revised / extended \\
\addlinespace[2pt]
8 & Belief, incentives, and mind-reading & Riskier strategies when safe payoffs fail; shared theory of mind aids coordination (Weber et al. 2004; Westby and Riedl 2023) & Asymmetric incentives (AI bears no stakes; humans under-rely under high stakes); coordination needs cross-type theory of mind, not anthropomorphism; group-level alignment does not follow from individual-level alignment & Gonzalez et al. 2026; Gupta et al. 2025; Keshmirian et al. 2025; Ashery et al. 2025; Hemmatian et al. in press & Revised \\
\addlinespace[2pt]
9 & Diversity over time & Conformity erodes diversity (Salganik et al. 2006) & AI homogenizes; hybrids sustain diversity longest & Shiiku et al. 2025; Burton et al. 2024; Doshi and Hauser 2024 & Revised \\
\addlinespace[2pt]
\bottomrule
\end{tabularx}
\end{sidewaystable}

The exploration-exploitation logic itself is robust across the dimensions and network types: For human, AI, and hybrid groups, structures that speed the spread of a current best answer trade away the diversity that hard problems need. But the trade-off now also turns on who sits where and what kinds of links exist, not only on the agent-agnostic network architecture, and this is what produces the unique hub, clustering, and annealing effects noted above. Efficiency versus redundancy, a minor key in the human literature, in turn, becomes a governing tension in hybrid systems, because the human-AI channel is lossy enough that the redundancy a homogeneous designer would prune is what keeps a hybrid group accurate.

\section{Research, Methods, and Knowledge Development}

So far, we have assembled the parameter space that the human and AI traditions established, re-expressed it through the memory, attention, reasoning (MAR) lens, and read off where hybrid networks behave similarly to homogeneous ones rather than diverge. While some assertions have empirical support from recent hybrid studies, many are educated, falsifiable conjectures. This section sketches how a research program could test the predictions laid out in Tables 4 and 5.

The parameter toolbox now applied to hybrid systems (Table 5) and the structure catalog (Table 4) together define a design space. Every conjecture takes the form that a parameter doing one thing in homogeneous networks does another in hybrid ones. Therefore, the natural experimental approach is to hold the parameter fixed and compare human-only, AI-only, and hybrid conditions against matched appropriate baselines. Then, manipulations of node and link types recover the headline conjectures. Varying the type of agent at a fixed network position tests the human-versus-AI hub claim (no. 1). Changing the share of AI nodes traces the predicted inverted-U performance in group composition (no. 2). Adjusting the number of agents and the depth of delegation probes the group size diminishing returns and verification-bottleneck predictions (no. 3). Rewiring to raise or lower local density tests whether the tighter clustering of AI subgroups amplifies cooperation and suppresses exploration at once (no. 4). Injecting a structural dissenter probes the proposed remedy for sycophancy when a group's learning strategy sits too close to its topology (no. 5). Introducing human-paced checkpoints in a hybrid network can test the effectiveness of hybrid paradigms that sequence exploration and exploitation (no. 6). Varying the capacity and noise of the human-AI link (e.g., by constraining message length, perturbing translation or adding prompt engineering instructions) tests whether collective performance is bound by the lowest-capacity cut of the network (no. 7). Finally, changing stakes and payoff asymmetry across agent types tests the predicted incentive mismatch and cross-type theory of mind effects (no. 8), while seeding groups with more or less heterogeneous agents tests the diversity conjecture (no. 9).

Each manipulation has a paradigm ready to implement it. For example, varying agent type at a fixed position (no. 1) is the natural use of a controlled co-creation testbed (e.g., Hemmatian et al., in press), which can swap a human or an AI partner into the same role under matched conditions and against the non-interactive baselines that separate interaction from raw agent ability. Changing the share and placement of AI nodes (e.g., no. 2) is what experimental social networks were built for, since participants occupy the nodes of a graph whose composition and wiring the experimenter controls and into which agents can be embedded at chosen positions (Shirado and Christakis 2017). Iterated-transmission versions of the same design carry it across generations to reach the temporal conjectures about dynamics and diversity over time (nos. 6 and 9; Shiiku et al. 2025). Varying the capacity and noise of the human-AI link (no. 7) is most easily done in LLM agent-based simulations and in benchmarks organized around information asymmetry, where message length, translation, and visibility can be set precisely and at scales no laboratory reaches (Liu et al. 2024; Han et al. 2026). Finally, the error decorrelation condition behind the ensemble performance conjecture can be tested directly by re-running aggregation studies with human and AI error profiles deliberately varied (Zöller et al. 2025).

Throughout, scalable homogeneous observation can complement hybrid manipulations: decades of human social-network corpora on one side (Lazer et al. 2009), and on the other AI-only platforms such as Moltbook, where tens of thousands of agents interact with one another in public (Zhang et al. 2026). Comparing the two is a way to characterize how the node and edge attributes of humans and of AI agents differ in the wild. This is precisely the input a hybrid theory needs, since better-specified boundary cases yield sharper conjectures about hybrid setups.

Regardless of the study design, testing the conjectures requires measuring more than outcomes. Accuracy, originality, and the diversity of a group's solutions capture what a collective produces, but the hybrid claims are mostly about process, so the quantities of interest also include transmission fidelity, the rate of consensus, the correlation of errors across agents, and the calibration of trust, the process measures that reveal fragility a raw accuracy score conceals (Gonzalez et al. 2026). Thankfully, the storied literature on human networks supplies many of the needed formalisms (Hills 2025). The information-theoretic framing of §4 provides a complementary quantitative vocabulary: channel capacity and noise for individual links, conductance for whole networks, and the redundancy of messages relative to their information content, each of which can be estimated and related to collective hybrid performance (Ayaso et al. 2010; Mani et al. 2021).

Two priorities should guide future research. The first is shared infrastructure: common testbeds, benchmarks, and reporting standards would let results accumulate across the cognitive-science, multi-agent-systems, and human-AI-teaming communities whose findings this chapter has described, rather than re-fragmenting along old disciplinary lines. The second is a caution about scope. Much of the hybrid evidence to date comes from small, short, laboratory, or prototype studies, and its generalization to large, long-lived, and heterogeneous real-world systems remains uncertain (Gonzalez et al. 2026). The most valuable future work will therefore be longitudinal and field-based, following hybrid collectives as they adapt, as a cognitive systems perspective insists they will, and testing whether the structural effects catalogued here persist once the agents on both sides of the interface change one another.

\section{Implications and Applications}

Our synthesis offers a clear set of structural levers that impact real-world downstream uses of hybrid systems. The \textit{team composition} is one: because a human in the minority loses trust and shared understanding (Gonzalez et al. 2026), and the effect of adding AI is non-monotone, the ratio of human to AI agents is a decision to make deliberately rather than leave to network economy considerations or chance. The placement of people is another structural design feature. The analysis of \textit{interface roles} implies that a human should be positioned according to the failure the system needs to prevent the most: as a gatekeeper where the danger is a flood of low-quality or low-diversity input, and as a supervisor where the danger is undetected error. The definition of the \textit{AI's objective} is a third. For instance, since an agent that controls a group's edges or attention imposes its goal on the whole network, a poorly specified or drifting objective is not a local bug but a systemic one (Dellermann et al. 2019; Gonzalez et al. 2026).

The risks that governance must address are the shadow side of this same structure, and they surface at every level of the network. At the level of the node, \textit{visible }\textit{identity }is the first concern: because a small fraction of well-placed agents can move a whole collective (Tsvetkova et al. 2024; Han et al. 2026), a covert AI participant is a structural hazard, which makes disclosure of which nodes are machines a precondition for trust and consequently collective performance. At the level of interaction, \textit{unintentional coordination }can happen: pricing algorithms left to interact can learn to sustain supra-competitive prices without ever communicating, a collusion invisible to a regulator watching only for an explicit agreement (Calvano et al. 2020). At the human-AI interface, a key danger is \textit{misplaced accountability}. For example, a person positioned at the output boundary merely to sign off can make them a moral crumple zone, absorbing blame for failures they had little power to prevent while the system's behavior is left unchanged (Elish 2019). At the level of the whole ecosystem, the \textit{homogenizing tendency} of AI collectives, compounded when models are retrained on their own output, threatens a slow collapse of diversity that no single network would notice but all would suffer (Shumailov et al. 2024; Han et al. 2026). Because such biases, conventions, and misalignments form at the level of the collective rather than the individual and can flip once a committed minority reaches a critical mass, alignment and oversight must themselves be assessed at the group level (Ashery et al. 2025; Keshmirian et al. 2025). The pattern across these levels is the same: the very feature that gives a hybrid network its power, the leverage of a few nodes, the ease of coordination, the human kept in the loop, the recycling of content, can also be the source of its characteristic failures.

But the rewards are worth the risks. Complementarity is real where it is engineered for: hybrid collectives outperform either part when their errors are diverse and their reliance is calibrated (Zöller et al. 2025), and the structural remedies catalogued here, a \textit{dissenting node} against premature consensus, \textit{redundancy} across a lossy interface, and \textit{annealing}-like pacing against runaway convergence, can be baked into networks. The preservation of the right level of diversity is the clearest prize: because humans retain idiosyncrasies that AI collectives shed, a well-composed hybrid can sustain the exploration that pure-AI systems lose over time (Shiiku et al. 2025; Doshi and Hauser 2024), which matters most for the open-ended, difficult problems hybrid intelligence is best-suited to address.

Two cautions temper our summary. First, none of the described effects are unconditional; the recurring lesson of the chapter is that a structure good for one task can be bad for another. Second, the hybrid evidence remains new, drawn largely from small and short-lived studies. Its transfer to large, durable, real-world systems is still to be demonstrated (Gonzalez et al. 2026). The implications above are therefore best read as design hypotheses, to be tested in the program of §5 before they are trusted in deployment.

\section{Relation to Other Topics in the Handbook}

This chapter supplies an integrated, bounded-cognition network framework for understanding hybrid collectives that several neighboring chapters approach from other directions. It is most directly complementary to treatments of hybrid teams as collective-intelligence systems (Singh, this volume) and to the behavioural sociology of humans and machines (Yasseri, this volume). Its account of communication and deliberation connects to the design of systems for large-scale collective deliberation (Pescetelli, this volume), and its treatment of the human node complements the social-cognition perspective on the human component of hybrid intelligence (Rydén and Svendsen, this volume), as well as the comparative conceptual framework for representation and coordination (Williams, this volume).

The implications of §6 link the chapter to the handbook's treatment of governance and organization, including hybrid authority and accountability (Tuncalp, this volume), the governed transition to hybrid intelligence (Polat, this volume), and the design of organizational hybrid systems (Liu, this volume). For each, the structures catalogued here help identify where in a network the decisive design and oversight choices lie. More broadly, by situating hybrid intelligence as the heterogeneous case of collective intelligence, the chapter offers the foundational sections a bridge between the established science of human and machine collectives and the new questions this handbook poses.

\section{Conclusion and Outlook}

We have argued that hybrid intelligence is best understood not as a new phenomenon but as the heterogeneous case of collective intelligence, and that the tools to study it already exist, divided between a cognitive science of human groups and an engineering science of multi-agent AI. Read through a common lens of memory, attention, and reasoning capacities, the two literatures supply a shared parameter toolbox. Brought together, the parameters in turn reveal which of the homogeneous network findings survive when a network's nodes and edges are of mixed type, which must be revised, and which structures, native to hybrid systems, neither established research tradition had reason to deeply study. The two trade-offs that organize the established fields, exploration versus exploitation and efficiency versus redundancy, both persist into the hybrid case but are transformed by it: the first now turns on who occupies which position, and the second, a minor concern among people, becomes central in human-AI channels.

Over the next five to ten years, as agentic AI is woven ever more deeply into human groups, the structural choices catalogued here, the composition of teams, the placement of people, the design of the human-AI interface, and the pacing of convergence, will increasingly determine whether a hybrid collective amplifies human intelligence or quietly erodes it. Treating these as deliberate design variables, rather than as accidents of procurement, is the practical promise of a network science of hybrid cognition.

The broadest take-away is that a hybrid collective is not a weighted average of the humans and machines that compose it. Heterogeneity at the nodes and along the edges creates genuinely new structures and re-maps familiar effects, so that the behavior of the network must be designed, measured, and governed independently. The synthesis offered here is a step toward doing so with the rigor the problem deserves.

\section*{References}
\begingroup\footnotesize\RaggedRight
\setlength{\parindent}{0pt}\setlength{\parskip}{3pt}
\hangindent=1.4em\hangafter=1\noindent Akata E, Schulz L, Coda-Forno J, Oh SJ, Bethge M, Schulz E (2025) Playing repeated games with large language models. \textit{Nature Human Behaviour} 9:1380--1390. doi:10.1038/\allowbreak{}s41562-025-02172-y\par
\hangindent=1.4em\hangafter=1\noindent Acemoglu D, Ozdaglar A (2011) Opinion dynamics and learning in social networks. Dynamic Games and Applications 1(1):3--49. doi:10.1007/\allowbreak{}s13235-010-0004-1\par
\hangindent=1.4em\hangafter=1\noindent Almaatouq A, Noriega-Campero A, Alotaibi A, Krafft PM, Moussaïd M, Pentland A (2020) Adaptive social networks promote the wisdom of crowds. Proc Natl Acad Sci USA 117(21):11379--11386. doi:10.1073/\allowbreak{}pnas.1917687117\par
\hangindent=1.4em\hangafter=1\noindent Aral S, Van Alstyne M (2011) The diversity-bandwidth trade-off. Am J Sociol 117(1):90--171. doi:10.1086/\allowbreak{}661238\par
\hangindent=1.4em\hangafter=1\noindent Ashery AF, Aiello LM, Baronchelli A (2025) Emergent social conventions and collective bias in LLM populations. Sci Adv 11(20):eadu9368. doi:10.1126/\allowbreak{}sciadv.adu9368\par
\hangindent=1.4em\hangafter=1\noindent Ayaso O, Shah D, Dahleh MA (2010) Information-theoretic bounds for distributed computation over networks of point-to-point channels. IEEE Trans Inf Theory 56(12):6020--6039. doi:10.1109/\allowbreak{}TIT.2010.2080910\par
\hangindent=1.4em\hangafter=1\noindent Baltaji R, Hemmatian B, Varshney LR (2024) Conformity, confabulation, and impersonation: persona inconstancy in multi-agent LLM collaboration. In: Proceedings of the 2nd Workshop on Cross-Cultural Considerations in NLP (C3NLP @ ACL 2024). arXiv:2405.03862\par
\hangindent=1.4em\hangafter=1\noindent Barabási A-L, Albert R (1999) Emergence of scaling in random networks. Science 286(5439):509--512. doi:10.1126/\allowbreak{}science.286.5439.509\par
\hangindent=1.4em\hangafter=1\noindent Barkoczi D, Galesic M (2016) Social learning strategies modify the effect of network structure on group performance. Nat Commun 7:13109. doi:10.1038/\allowbreak{}ncomms13109\par
\hangindent=1.4em\hangafter=1\noindent Boyd R, Richerson PJ (2005) The origin and evolution of cultures. Oxford University Press, New York\par
\hangindent=1.4em\hangafter=1\noindent Brady WJ, Crockett MJ, Van Bavel JJ (2020) The MAD model of moral contagion: the role of motivation, attention, and design in the spread of moralized content online. Perspect Psychol Sci 15(4):978--1010. doi:10.1177/\allowbreak{}1745691620917336\par
\hangindent=1.4em\hangafter=1\noindent Burton JW, Lopez-Lopez E, Hechtlinger S, et al. (2024) How large language models can reshape collective intelligence. Nat Hum Behav 8(9):1643--1655. doi:10.1038/\allowbreak{}s41562-024-01959-9\par
\hangindent=1.4em\hangafter=1\noindent Calvano E, Calzolari G, Denicolò V, Pastorello S (2020) Artificial intelligence, algorithmic pricing, and collusion. Am Econ Rev 110(10):3267--3297. doi:10.1257/\allowbreak{}aer.20190623\par
\hangindent=1.4em\hangafter=1\noindent Campbell CM, Izquierdo EJ, Goldstone RL (2022) Partial copying and the role of diversity in social learning performance. Collect Intell 1(1):26339137221081849. doi:10.1177/\allowbreak{}26339137221081849\par
\hangindent=1.4em\hangafter=1\noindent Centola D (2018) How behavior spreads: the science of complex contagions. Princeton University Press, Princeton\par
\hangindent=1.4em\hangafter=1\noindent Centola D (2022) The network science of collective intelligence. Trends Cogn Sci 26(11):923--941. doi:10.1016/\allowbreak{}j.tics.2022.08.009\par
\hangindent=1.4em\hangafter=1\noindent Centola D, Baronchelli A (2015) The spontaneous emergence of conventions: an experimental study of cultural evolution. Proc Natl Acad Sci USA 112(7):1989--1994. doi:10.1073/\allowbreak{}pnas.1418838112\par
\hangindent=1.4em\hangafter=1\noindent Centola D, Becker J, Brackbill D, Baronchelli A (2018) Experimental evidence for tipping points in social convention. Science 360(6393):1116--1119. doi:10.1126/\allowbreak{}science.aas8827\par
\hangindent=1.4em\hangafter=1\noindent Charnov EL (1976) Optimal foraging, the marginal value theorem. Theor Popul Biol 9(2):129--136. doi:10.1016/\allowbreak{}0040-5809(76)90040-X\par
\hangindent=1.4em\hangafter=1\noindent Clark HH, Brennan SE (1991) Grounding in communication. In: Resnick LB, Levine JM, Teasley SD (eds) Perspectives on socially shared cognition. American Psychological Association, Washington, DC, pp 127--149\par
\hangindent=1.4em\hangafter=1\noindent Cui H, Yasseri T (2024) AI-enhanced collective intelligence. Patterns 5(11):101074. doi:10.1016/\allowbreak{}j.patter.2024.101074\par
\hangindent=1.4em\hangafter=1\noindent Dellermann D, Ebel P, Söllner M, Leimeister JM (2019) Hybrid intelligence. Bus Inf Syst Eng 61(5):637--643. doi:10.1007/\allowbreak{}s12599-019-00595-2\par
\hangindent=1.4em\hangafter=1\noindent Derex M, Boyd R (2016) Partial connectivity increases cultural accumulation within groups. Proc Natl Acad Sci USA 113(11):2982--2987. doi:10.1073/\allowbreak{}pnas.1518798113\par
\hangindent=1.4em\hangafter=1\noindent Doshi AR, Hauser OP (2024) Generative AI enhances individual creativity but reduces the collective diversity of novel content. Sci Adv 10(28):eadn5290. doi:10.1126/\allowbreak{}sciadv.adn5290\par
\hangindent=1.4em\hangafter=1\noindent Du Y, Li S, Torralba A, Tenenbaum JB, Mordatch I (2023) Improving factuality and reasoning in language models through multiagent debate. arXiv:2305.14325 (published in Proceedings of the 41st International Conference on Machine Learning, ICML 2024, PMLR 235:11733--11763)\par
\hangindent=1.4em\hangafter=1\noindent Easley D, Kleinberg J (2010) Networks, crowds, and markets: reasoning about a highly connected world. Cambridge University Press, Cambridge\par
\hangindent=1.4em\hangafter=1\noindent Elish MC (2019) Moral crumple zones: cautionary tales in human-robot interaction. Engag Sci Technol Soc 5:40--60. doi:10.17351/\allowbreak{}ests2019.260\par
\hangindent=1.4em\hangafter=1\noindent Golub B, Jackson MO (2010) Naïve learning in social networks and the wisdom of crowds. American Economic Journal: Microeconomics 2(1):112--149. doi:10.1257/\allowbreak{}mic.2.1.112\par
\hangindent=1.4em\hangafter=1\noindent Gonzalez C, Donahue K, Goldstein DG, Heidari H, Jalali MS, Schelble B, Singh A, Woolley AW (2026) Toward a science of human--AI teaming for decision making: a complementarity framework. PNAS Nexus 5(3):pgag030. doi:10.1093/\allowbreak{}pnasnexus/\allowbreak{}pgag030\par
\hangindent=1.4em\hangafter=1\noindent Grötschla F, Müller L, Tönshoff J, Galkin M, Perozzi B (2025) AgentsNet: coordination and collaborative reasoning in multi-agent LLMs. arXiv:2507.08616\par
\hangindent=1.4em\hangafter=1\noindent Guo T, Chen X, Wang Y, Chang R, Pei S, Chawla NV, Wiest O, Zhang X (2024) Large language model based multi-agents: a survey of progress and challenges. In: Proceedings of the 33rd International Joint Conference on Artificial Intelligence (IJCAI 2024). arXiv:2402.01680\par
\hangindent=1.4em\hangafter=1\noindent Gupta P, Nguyen TN, Gonzalez C, Woolley AW (2025) Fostering collective intelligence in human--AI collaboration: laying the groundwork for COHUMAIN. Top Cogn Sci 17(2):189--216. doi:10.1111/\allowbreak{}tops.12679\par
\hangindent=1.4em\hangafter=1\noindent Gupta P, Zhong Q, Yakura H, Eisenmann T, Rahwan I (2026) The role of social learning and collective norm formation in fostering cooperation in LLM multi-agent systems. In: Proceedings of the 25th International Conference on Autonomous Agents and Multiagent Systems (AAMAS 2026). arXiv:2510.14401\par
\hangindent=1.4em\hangafter=1\noindent Han TA, Leibo JZ, Lenaerts T, Rahwan I, Santos FP, Perc M, Capraro V (2026) Social physics in the age of artificial intelligence. arXiv:2603.16900\par
\hangindent=1.4em\hangafter=1\noindent Haupt A, Brynjolfsson E (2025) Position: AI should not be an imitation game---centaur evaluations. In: Proceedings of the 42nd International Conference on Machine Learning (ICML 2025), Position Paper Track\par
\hangindent=1.4em\hangafter=1\noindent Hemmatian B, Sloman SA (2020) Two systems for thinking with a community: outsourcing versus collaboration. In: Elqayam S, Douven I, Evans JSBT, Cruz N (eds) Logic and uncertainty in the human mind: a tribute to David E. Over. Routledge, New York, pp 102--115\par
\hangindent=1.4em\hangafter=1\noindent Hemmatian B, Keshmirian A, Ramamoorthy S, Jahadakbar M, Khuri-Reid E, Wang J, Hadjarab S, Veum S, Lin Y, Gupta P, Somaya D, Varshney LR (in press) Two-player alternate uses test: a controlled testbed for interactive human--AI and human--human co-creation. In: Proceedings of the 2026 ACM Conference on Creativity \& Cognition (C\&C '26)\par
\hangindent=1.4em\hangafter=1\noindent Hills TT (2025) Behavioral network science: language, mind, and society. Cambridge University Press, Cambridge. doi:10.1017/\allowbreak{}9781108883894\par
\hangindent=1.4em\hangafter=1\noindent Hong L, Page SE (2004) Groups of diverse problem solvers can outperform groups of high-ability problem solvers. Proc Natl Acad Sci USA 101(46):16385--16389. doi:10.1073/\allowbreak{}pnas.0403723101\par
\hangindent=1.4em\hangafter=1\noindent Huang M, Zhao Q, Yi X, Xie X (2025) On the dynamics of multi-agent LLM communities driven by value diversity. arXiv:2512.10665\par
\hangindent=1.4em\hangafter=1\noindent Keshmirian A, Baltaji R, Hemmatian B, Asghari H, Varshney LR (2025) Many LLMs are more utilitarian than one. In: Advances in Neural Information Processing Systems 38 (NeurIPS 2025). arXiv:2507.00814\par
\hangindent=1.4em\hangafter=1\noindent Kong W, et al. (2026) Multi-LLM systems exhibit robust semantic collapse. arXiv:2605.17193\par
\hangindent=1.4em\hangafter=1\noindent Kumarappan A, Mujoo A (2026) Not just RLHF: why alignment alone won't fix multi-agent sycophancy. arXiv:2605.12991\par
\hangindent=1.4em\hangafter=1\noindent Latora V, Marchiori M (2001) Efficient behavior of small-world networks. Phys Rev Lett 87(19):198701. doi:10.1103/\allowbreak{}PhysRevLett.87.198701\par
\hangindent=1.4em\hangafter=1\noindent Lazer D, Pentland A, Adamic L, Aral S, Barabási A-L, Brewer D, et al. (2009) Computational social science. Science 323(5915):721--723. doi:10.1126/\allowbreak{}science.1167742\par
\hangindent=1.4em\hangafter=1\noindent Leibo JZ, Zambaldi V, Lanctot M, Marecki J, Graepel T (2017) Multi-agent reinforcement learning in sequential social dilemmas. In: Proceedings of the 16th International Conference on Autonomous Agents and Multiagent Systems (AAMAS 2017), pp 464--473. arXiv:1702.03037\par
\hangindent=1.4em\hangafter=1\noindent Li G, Hammoud H, Itani H, Khizbullin D, Ghanem B (2023) CAMEL: communicative agents for ``mind'' exploration of large language model society. In: Advances in Neural Information Processing Systems 36 (NeurIPS 2023)\par
\hangindent=1.4em\hangafter=1\noindent Lion S, van Baalen M (2008) Self-structuring in spatial evolutionary ecology. Ecol Lett 11(3):277--295. doi:10.1111/\allowbreak{}j.1461-0248.2007.01132.x\par
\hangindent=1.4em\hangafter=1\noindent Liu W, Wang C, Wang Y, Xie Z, Qiu R, Dang Y, Du Z, Chen W, Yang C, Qian C (2024) Autonomous agents for collaborative task under information asymmetry. In: Advances in Neural Information Processing Systems 37 (NeurIPS 2024). arXiv:2406.14928\par
\hangindent=1.4em\hangafter=1\noindent Lorenz J, Rauhut H, Schweitzer F, Helbing D (2011) How social influence can undermine the wisdom of crowd effect. Proc Natl Acad Sci USA 108(22):9020--9025. doi:10.1073/\allowbreak{}pnas.1008636108\par
\hangindent=1.4em\hangafter=1\noindent Madsen JK, Bailey RM, Pilditch TD (2018) Large networks of rational agents form persistent echo chambers. Scientific Reports 8:12391. doi:10.1038/\allowbreak{}s41598-018-25558-7\par
\hangindent=1.4em\hangafter=1\noindent Mani A, Varshney LR, Pentland A (2021) Quantization games on social networks and language evolution. IEEE Trans Signal Process 69:3922--3934. arXiv:2006.00584\par
\hangindent=1.4em\hangafter=1\noindent March JG (1991) Exploration and exploitation in organizational learning. Organ Sci 2(1):71--87. doi:10.1287/\allowbreak{}orsc.2.1.71\par
\hangindent=1.4em\hangafter=1\noindent Mao A, Mason W, Suri S, Watts DJ (2016) An experimental study of team size and performance on a complex task. PLoS ONE 11(4):e0153048. doi:10.1371/\allowbreak{}journal.pone.0153048\par
\hangindent=1.4em\hangafter=1\noindent Mason WA, Jones A, Goldstone RL (2008) Propagation of innovations in networked groups. J Exp Psychol Gen 137(3):422--433. doi:10.1037/\allowbreak{}a0012798\par
\hangindent=1.4em\hangafter=1\noindent Mason W, Watts DJ (2012) Collaborative learning in networks. Proc Natl Acad Sci USA 109(3):764--769. doi:10.1073/\allowbreak{}pnas.1110069108\par
\hangindent=1.4em\hangafter=1\noindent McKee KR, Tacchetti A, Bakker MA, Balaguer J, Campbell-Gillingham L, Everett R, Botvinick M (2023) Scaffolding cooperation in human groups with deep reinforcement learning. Nat Hum Behav 7(10):1787--1796. doi:10.1038/\allowbreak{}s41562-023-01686-7\par
\hangindent=1.4em\hangafter=1\noindent Mehlhorn K, Newell BR, Todd PM, Lee MD, Morgan K, Braithwaite VA, Hausmann D, Fiedler K, Gonzalez C (2015) Unpacking the exploration--exploitation tradeoff: a synthesis of human and animal literatures. Decision 2(3):191--215. doi:10.1037/\allowbreak{}dec0000033\par
\hangindent=1.4em\hangafter=1\noindent Newman MEJ (2018) Networks, 2nd edn. Oxford University Press, Oxford\par
\hangindent=1.4em\hangafter=1\noindent O'Neill T, McNeese N, Barron A, Schelble B (2022) Human--autonomy teaming: a review and analysis of the empirical literature. Hum Factors 64(5):904--938. doi:10.1177/\allowbreak{}0018720820960865\par
\hangindent=1.4em\hangafter=1\noindent Park JS, O'Brien JC, Cai CJ, Morris MR, Liang P, Bernstein MS (2023) Generative agents: interactive simulacra of human behavior. In: Proceedings of the 36th Annual ACM Symposium on User Interface Software and Technology (UIST '23). doi:10.1145/\allowbreak{}3586183.3606763\par
\hangindent=1.4em\hangafter=1\noindent Perc M, Jordan JJ, Rand DG, Wang Z, Boccaletti S, Szolnoki A (2017) Statistical physics of human cooperation. Phys Rep 687:1--51. doi:10.1016/\allowbreak{}j.physrep.2017.05.004\par
\hangindent=1.4em\hangafter=1\noindent Perez J, Léger C, Ovando-Tellez M, Foulon C, Dussauld J, Oudeyer P-Y, Moulin-Frier C (2024) Cultural evolution in populations of large language models. arXiv:2403.08882\par
\hangindent=1.4em\hangafter=1\noindent Riedl C, Savage S, Zvelebilova J (2024) AI's social forcefield: reshaping distributed cognition in human-AI teams. arXiv:2407.17489\par
\hangindent=1.4em\hangafter=1\noindent Salganik MJ, Dodds PS, Watts DJ (2006) Experimental study of inequality and unpredictability in an artificial cultural market. Science 311(5762):854--856. doi:10.1126/\allowbreak{}science.1121066\par
\hangindent=1.4em\hangafter=1\noindent Seeber I, Bittner E, Briggs RO, de Vreede T, de Vreede GJ, Elkins A, et al. (2020) Machines as teammates: a research agenda on AI in team collaboration. Inf Manag 57(2):103174. doi:10.1016/\allowbreak{}j.im.2019.103174\par
\hangindent=1.4em\hangafter=1\noindent Shen X, Liu Y, Dai Y, Wang Y, Miao R, Tan Y, Pan S, Wang X (2025) Understanding the information propagation effects of communication topologies in LLM-based multi-agent systems. arXiv:2505.23352\par
\hangindent=1.4em\hangafter=1\noindent Shiiku S, Marjieh R, Anglada-Tort M, Jacoby N (2025) The dynamics of collective creativity in human-AI hybrid societies. In: Proceedings of the 47th Annual Conference of the Cognitive Science Society (CogSci 2025). arXiv:2502.17962\par
\hangindent=1.4em\hangafter=1\noindent Shirado H, Christakis NA (2017) Locally noisy autonomous agents improve global human coordination in network experiments. Nature 545(7654):370--374. doi:10.1038/\allowbreak{}nature22332\par
\hangindent=1.4em\hangafter=1\noindent Shumailov I, Shumaylov Z, Zhao Y, Papernot N, Anderson R, Gal Y (2024) AI models collapse when trained on recursively generated data. Nature 631(8022):755--759. doi:10.1038/\allowbreak{}s41586-024-07566-y\par
\hangindent=1.4em\hangafter=1\noindent Steiner ID (1972) Group process and productivity. Academic Press, New York\par
\hangindent=1.4em\hangafter=1\noindent Sutton RS, Barto AG (2018) Reinforcement learning: an introduction, 2nd edn. MIT Press, Cambridge, MA\par
\hangindent=1.4em\hangafter=1\noindent Tsvetkova M, Yasseri T, Pescetelli N, Werner T (2024) A new sociology of humans and machines. Nat Hum Behav 8(10):1864--1876. doi:10.1038/\allowbreak{}s41562-024-02001-8\par
\hangindent=1.4em\hangafter=1\noindent Ueshima A, Jones MI, Christakis NA (2024) Simple autonomous agents can enhance creative semantic discovery by human groups. Nat Commun 15(1):5212. doi:10.1038/\allowbreak{}s41467-024-49528-y\par
\hangindent=1.4em\hangafter=1\noindent Watts DJ, Strogatz SH (1998) Collective dynamics of `small-world' networks. Nature 393(6684):440--442. doi:10.1038/\allowbreak{}30918\par
\hangindent=1.4em\hangafter=1\noindent Weber EU, Shafir S, Blais A-R (2004) Predicting risk sensitivity in humans and lower animals: risk as variance or coefficient of variation. Psychol Rev 111(2):430--445. doi:10.1037/\allowbreak{}0033-295X.111.2.430\par
\hangindent=1.4em\hangafter=1\noindent Westby S, Riedl C (2023) Collective intelligence in human-AI teams: a Bayesian theory of mind approach. Proc AAAI Conf Artif Intell 37(5):6119--6127. doi:10.1609/\allowbreak{}aaai.v37i5.25755\par
\hangindent=1.4em\hangafter=1\noindent Wilcox RR, Hemmatian B, Varshney LR, Barbey AK (2026) The network architecture of general intelligence in the human connectome. Nat Commun 17. doi:10.1038/\allowbreak{}s41467-026-68698-5\par
\hangindent=1.4em\hangafter=1\noindent Woolley AW, Chabris CF, Pentland A, Hashmi N, Malone TW (2010) Evidence for a collective intelligence factor in the performance of human groups. Science 330(6004):686--688. doi:10.1126/\allowbreak{}science.1193147\par
\hangindent=1.4em\hangafter=1\noindent Wynn A, Satija H, Hadfield G (2025) Talk isn't always cheap: understanding failure modes in multi-agent debate. arXiv:2509.05396\par
\hangindent=1.4em\hangafter=1\noindent Wooldridge M (2009) An introduction to multiagent systems, 2nd edn. Wiley, Chichester\par
\hangindent=1.4em\hangafter=1\noindent Zamfirescu-Pereira JD, Wong RY, Hartmann B, Yang Q (2023) Why Johnny can't prompt: how non-AI experts try (and fail) to design LLM prompts. In: Proceedings of the 2023 CHI Conference on Human Factors in Computing Systems (CHI '23). doi:10.1145/\allowbreak{}3544548.3581388\par
\hangindent=1.4em\hangafter=1\noindent Zhang Y, Mei K, Liu M, Wang J, Metaxas DN, Wang X, Hamm J, Ge Y (2026) Agents in the wild: safety, society, and the illusion of sociality on Moltbook. arXiv:2602.13284\par
\hangindent=1.4em\hangafter=1\noindent Zöller N, et al. (2025) Human--AI collectives most accurately diagnose clinical vignettes. Proc Natl Acad Sci USA 122(24):e2426153122. doi:10.1073/\allowbreak{}pnas.2426153122\par
\hangindent=1.4em\hangafter=1\noindent Zhang G, Yue Y, Li Z, Yun S, Wan G, Wang K, Cheng D, Yu J, Chen T (2025). Cut the crap: An economical communication pipeline for llm-based multi-agent systems. In: International Conference on Learning Representations 2025 May 1 (Vol. 2025, pp. 75389-75428).\par
\hangindent=1.4em\hangafter=1\noindent Nisioti E, Risi S, Momennejad I, Oudeyer PY, Moulin-Frier C (2024). Collective innovation in groups of large language models. arXiv:2407.05377.\par
\hangindent=1.4em\hangafter=1\noindent Qian C, Xie Z, Wang Y, Liu W, Zhu K, Xia H, Dang Y, Du Z, Chen W, Yang C, Liu Z (2025). Scaling large language model-based multi-agent collaboration. In: International Conference on Learning Representations 2025 May 1 (Vol. 2025, pp. 41488-41505).\par
\hangindent=1.4em\hangafter=1\noindent Kalai AT, Nachum O, Vempala SS, Zhang E (2025). Why language models hallucinate. arXiv preprint arXiv:2509.04664\par
\hangindent=1.4em\hangafter=1\noindent Chuang, Y., Goyal, A., Harlalka, N., Suresh, S., Hawkins, R., Yang, S., Shah, D., Hu, J., \& Rogers, T.T. (2023). Simulating Opinion Dynamics with Networks of LLM-based Agents. NAACL-HLT.\par
\endgroup
\end{document}